\documentclass[amsmath,amssymb,12pt,aps]{revtex4-2}
\usepackage[utf8]{inputenc}
\usepackage{graphicx}

\usepackage{natbib}
\usepackage{wrapfig}
\usepackage{xcolor}
\usepackage{dcolumn}
\usepackage[english]{babel}
\usepackage[normalem]{ulem}
\usepackage[hidelinks]{hyperref}

\newcommand\msout{\bgroup\markoverwith{\textcolor{red}{\rule[0.5ex]{1pt}{1pt}}}\ULon} 
\newcommand\leoout{\bgroup\markoverwith{\textcolor{blue}{\rule[0.5ex]{1pt}{1pt}}}\ULon} 

\begin{document}
	
\title{Modules as effective nodes in coarse-grained networks of Kuramoto oscillators}

\author{Leonardo L. Bosnardo}
\author{Marcus A. M. de Aguiar}
\affiliation{Instituto de F\'isica `Gleb Wataghin', Universidade Estadual de Campinas, Unicamp 13083-970, Campinas, SP, Brazil}

\begin{abstract}

    Most real-world networks exhibit a significant degree of modularity. Understanding the effects of such topology on dynamical processes is pivotal for advances in social and natural sciences. In this work we consider the dynamics of Kuramoto oscillators on modular networks and propose a simple coarse-graining procedure where modules are replaced by effective single oscillators. The method is inspired by EEG measurements, where very large groups of neurons under each electrode are interpreted as single nodes in a correlation network. We expose the interplay between intra-module and inter-module coupling strengths in keeping the coarse-graining process meaningful. We show that, when modules are well synchronized, the phase transition from asynchronous to synchronous motion in networks with 2 and 3 modules is very well described by their respective reduced systems, regardless of the network structure connecting the modules. Applications of the method to real networks with small modularity coefficients reveals that the approximation is also very accurate if oscillators in each module are identical. The method reproduces global synchronization patterns despite the low synchronizability of some modules, possibly allowing for the inference of the mean synchrony of each module when individual dynamics are not known.
    

\end{abstract}

\maketitle

\section{Introduction}
\label{intro}

Real world systems that exhibit synchronization are commonly contained in large networks of non-linear oscillators, such as neuronal networks \cite{cumin2007generalising,bhowmik2012well,ferrari2015phase,reis2021bursting} and power grids \cite{filatrella2008analysis,motter2013spontaneous,Nishikawa_2015,molnar2021asymmetry}. In these cases, measuring the individual state of each node is challenging, and corse-grained procedures are often employed \cite{meshulam2019coarse}. For example, measurements of brain activity using functional magnetic resonance imaging (fMRI) \cite{biswal1995functional,xiong1999interregional}, near-infrared spectroscopy (NIRS) \cite{mesquita2010resting} and electroencephalograms (EEG) \cite{mormann2000mean}, assess the average behavior of large regions of the brain, instead of capturing the oscillations of individual neurons. Analysis of correlations between such regions is used to infer patterns related to resting states or the performance of specific tasks. Neuronal diseases such as autism, Parkinson, schizophrenia, and epilepsy, for example, are commonly associated with abnormal modular synchronization \cite{uhlhaas2006neural,dinstein2011disrupted,hammond2007pathological,hong2004evoked,jiruska2013synchronization,mormann2000mean}.

Motivated by the difficulty of measuring and simulating large networks of oscillators, and by current EEG techniques, that always measures the average behavior of massive number of neurons in each electrode, we develop an approximation where groups of nodes are replaced by a single effective node, drastically reducing the size of the system. The theory is developed for modular networks, since modules form natural sub-groups of nodes, and we use the Kuramoto model as underlying dynamics.

In the Kuramoto model each oscillator is characterized by a single phase $\theta_i$ and its dynamics depends on other oscillators according to the equation \cite{kuramoto1975self,kuramoto1984chemical,acebron2005kuramoto,pikovsky2001universal} 
\begin{equation}
    \dot{\theta_{i}} = \omega_{i} + \frac{\lambda}{N} \sum_{j=1}^{N} \sin{(\theta_{j}-\theta_{i})}
\label{ku}
\end{equation}
where $N$ is the total number of oscillators and the natural angular velocity $\omega_i$ is usually chosen from a symmetric and unimodal distribution $g(\omega)$ centered at $\omega_0$. Each oscillator interacts with all the others according to their phase difference and the interactions are modulated by a global coupling strength $\lambda$. In the limit $N \rightarrow \infty$, the system undergoes a continuous phase transition from disordered motion to synchronization at $\lambda_\text{c} = 2/(\pi g(\omega_0))$. The phase transition can be characterized by the complex number
\begin{equation}
    z = r e^{i\psi} = \frac{1}{N} \sum_{j=1}^{N} e^{i\theta_{j}}
\label{z}
\end{equation}
representing the average of all phases. The module of $z$ is the order parameter of the transition, going from $r=0$, when motion is disordered, to $r=1$ for perfect synchronization.

The original assumption that each oscillator interacts with all the others is a simplifying approximation that fails in many real systems, such as neurons in the brain, that are grouped in well-defined regions \cite{hagmann2008mapping,honey2009predicting,winding2023connectome,white1986structure,chen2008revealing}, fireflies \cite{moiseff2010firefly,ermentrout1991adaptive,buck1976synchronous}, that only interact with close neighbors, and traffic routing, that depends critically on synchronization between neighbors \cite{chen2023traffic}. The extension of the Kuramoto model to networks \cite{arenas2008synchronization,odor2019critical,villegas2014frustrated,villegas2016complex,schmidt2015kuramoto} describing the set of possible pairwise interactions, revealed that the topology of connections has a large effect on the synchronization properties of the system, leading, for example, to frequency \cite{ocampo2025frequency} and explosive synchronization \cite{gomez2011explosive,zhang2013explosive,zhang2015explosive,ji2013cluster,chen2023traffic}. Another important aspect of the Kuramoto dynamics is how network properties, such as heterogeneity and symmetries, affect the paths to synchronization \cite{gomez2007paths}, which might involve the formation of clusters of synchronized oscillators resembling modules \cite{pecora2014cluster}. 

Here we propose an approximation to the dynamics of modular networks where each module is replaced by a single effective node. We compute the Kuramoto order parameter and study the conditions for the phase transition from disordered to synchronized state to be well represented by the reduced system. We find that such an approximation can be very good for a range of internal and inter-module coupling strengths. Moreover, we find that, under these conditions, the results are largely independent of the network of connections between modules, that can be ignored and replaced by a single link between the nodes representing the modules. This approach can be viewed as an alternative to other techniques based on the renormalization group that may rely on hyperbolic spaces \cite{garcia2018multiscale}, Laplacian matrices \cite{villegas2023laplacian,loures2023laplacian,schmidt2024spectral} and machine learning \cite{zhang2025coarse}, which have been used to describe large networks in terms of fewer supernodes that retain the basic properties of the original system. One of the advantages of our method is that it can be applied to finite networks. It requires, however, that the modules are well defined, in a way that will become clear in the next sections. We will show examples with synthetic networks and also numerical simulations for Zachary's Karate club social network (two modules) and the \textit{C. Elegans} gap junctions neural network (divided into three, five and ten modules).

This paper is organized as follows: in sections \ref{modular} and \ref{order parameter} we develop our coarse graining procedure for modular networks and in sections \ref{two} and \ref{three} we apply it to the simple cases of two and three modules respectively, using synthetic networks. In section \ref{real} we apply the coarse graining method to Zachary's Karate club and \textit{C. Elegans} gap junctions networks. Finally in section \ref{conclusions} we discuss our findings.

\section{Coarse graining for modular networks}
\label{modular}

Networks can be described by an adjacency matrix $A$ containing information about the coupling between pairs of nodes. Here we consider only undirected networks, where $A_{ij}=A_{ji}=1$ if nodes $i$ and $j$ interact and $A_{ij}=0$ otherwise. The extension of the Kuramoto model to networks is given by
\begin{equation}
    \dot{\theta_{i}} = \omega_{i} + \sum_{j=1}^{N} \lambda_{ij} A_{ij} \sin{(\theta_{j}-\theta_{i})}
\label{kunet}
\end{equation}
where $\lambda_{ij}$ defines the coupling strength between nodes $i$ and $j$.

It is usual in network theory to define modules as groups of nodes that are more densely connected to each other than to the rest of the network. Modular structures can be constructed by either decreasing the number of connections between nodes of different modules or by distinguishing between intra-module and inter-module connection strengths.

In order to test our coarse graining procedure we will first apply it to synthetic modular networks. These networks are construct as follows: modules are indexed by $\sigma=1, 2, \dots, s$ and contain $N_\sigma$ nodes, or oscillators, with $N=\sum_{\sigma=1}^s N_\sigma$. We call $A^{\sigma\sigma'}$ the block of the adjacency matrix between modules $\sigma$ and $\sigma'$. Nodes belonging to different modules are connected with probability $p$ and strength $\lambda_{\sigma\sigma'} = \lambda_{\sigma'\sigma}$. Therefore, for $p=0$ the network has $s$ disjoint modules. Similarly, oscillators belonging to the same module are connected according to $A^{\sigma\sigma}$ with strength $\lambda_{\sigma\sigma}$. We will say more about the structure of $A^{\sigma\sigma}$ shortly.

For this type of network Eq.(\ref{kunet}) can be written as 
\begin{equation}
    \dot{\theta}_{\sigma,i} = \omega_{\sigma,i} + \sum_{\sigma'=1}^{s} \frac{\lambda_{\sigma\sigma'}}{\langle k \rangle_{\sigma\sigma'}} \sum_{j=1}^{N_{\sigma'}} A_{ij}^{\sigma \sigma'}\sin{(\theta_{\sigma',j}-\theta_{\sigma,i})}
\label{multi pop II/ku}
\end{equation}
where $i=1,\dots,N_\sigma$ refers to an oscillator in module $\sigma$. Note that for a single fully connected module, the Kuramoto model, Eq. \eqref{ku} has a normalization $1/N$ on the interaction term. Now that each oscillator has a varied number of connections, we changed $N$ to $\langle k \rangle_{\sigma\sigma'}= \sum_i k_{i\sigma\sigma'} /N_\sigma$, the average number of connections that an oscillator from module $\sigma$ has with the whole module $\sigma'$, with $k_{i\sigma\sigma'} = \sum_j A^{\sigma\sigma'}_{ij}$. In the limit of a fully connected network we recover $\langle k \rangle_{\sigma\sigma'}\rightarrow N_{\sigma'}$. Note also that $\langle k \rangle_{\sigma\sigma'} N_{\sigma}$ is the total number of connections between modules.

Summing Eq. \eqref{multi pop II/ku} over $i$ and dividing by $N_{\sigma}$ we get 
\begin{equation}
    \langle\dot{\theta}_\sigma\rangle = \langle\omega_\sigma\rangle + \sum_{\sigma'=1}^{s} \frac{\lambda_{\sigma\sigma'}}{\langle k \rangle_{\sigma\sigma'} N_\sigma} \sum_{i=1}^{N_\sigma} \sum_{j=1}^{N_{\sigma'}} A_{ij}^{\sigma \sigma'}\sin{(\theta_{\sigma',j}-\theta_{\sigma,i})}
\label{multi pop II/ku summed i}
\end{equation}
where $\langle\dot{\theta}_\sigma\rangle$ and $\langle\omega_\sigma\rangle$ denote the average velocity and natural frequency of module $\sigma$.

The average velocity is a variable regarding the module. If we were able to write Eq. \eqref{multi pop II/ku summed i} solely in terms of average module variables we could simplify the dynamics to $s$ equations, drastically reducing the dimensionality of the problem. This can be achieved in the special case where each and every module has a large enough internal synchrony such that the phases of oscillators belonging to the same module are about equal. In other words, if $\lambda_{\sigma\sigma}$ is large enough for $\theta_{\sigma,i}\approx\langle\theta_{\sigma}\rangle$, then we can lose the symbol $\langle\,.\,\rangle$ to arrive at
\begin{equation}
    \dot{\theta}_\sigma = \omega_\sigma + \sum_{\sigma'=1}^{s}\lambda_{\sigma\sigma'} \sin{(\theta_{\sigma'}-\theta_\sigma)}
\label{multi pop II/ku sigma indexed}
\end{equation}
which is the Kuramoto model in a fully connected network, independent of the inter-module adjacency matrix $A^{\sigma\sigma'}$. Note that, in this coarse-graining limit, the intra-module connection terms in Eq. \eqref{multi pop II/ku summed i} vanish due to the coherent phase of its oscillators. Therefore, the topological structure of the adjacency matrix $A$ becomes irrelevant and could assume any form. Since the coefficients $\lambda_{\sigma\sigma'}$ do not need to be all equal, this constitutes an asymmetric network: the strength of the interaction that module $\sigma$ has with module $\sigma'$ is not necessarily the same as $\sigma'$ has with $\sigma$. 

Eq. \eqref{multi pop II/ku sigma indexed} shows that, under model \eqref{multi pop II/ku}, the dynamics of $s$ modules is qualitatively the same as that of $s$ oscillators if high synchrony within modules is satisfied. This simple observation summarizes the analytical basis for the coarse-graining process. In the next sections we will quantify the validity of this result by simulating model \eqref{multi pop II/ku} for $s=2$ and 3 to compare with the dynamics of 2 and 3 oscillators, respectively. Only symmetric cases will be treated, i.e, $\lambda_{\sigma\sigma}\equiv\lambda_{\text{in}}$ $\forall \sigma$ and $\lambda_{\sigma\sigma'}=\lambda_{\sigma'\sigma}\equiv\lambda/s$ $\forall \sigma\neq\sigma'$. We shall see that Eq. \eqref{multi pop II/ku sigma indexed} alone does not reproduce the behavior of the order parameter of the full modular network. However, when weights related to the degree of internal synchrony and relative module size are introduced to renormalize the contribution of the coarse grained nodes, very good agreement is observed.

Finally, we note that Eq. \eqref{multi pop II/ku sigma indexed} is also valid for arbitrary modular networks, not just the synthetic types we described, as long as the requirement of high synchrony within modules is satisfied. Numerical simulations in section \ref{real}, however, show that in real networks not all modules synchronize easily. Even in these cases, the coarse-graining procedure works very well with minor adjustments that take that into account this local lack of synchrony.

\section{Order parameter}
\label{order parameter}

In this section we propose an expression to calculate the order parameter of the coarse-grained network that matches, within our approximation, that of the original network. We will actually estimate upper and lower bounds for $r$ considering the limits that maintain the coarse-graining method valid. We start from Eq. \eqref{z}, that represents the global order parameter for the full network, and re-write it as
\begin{equation}
    z = \frac{1}{N} \sum_{\sigma=1}^{s} \sum_{j=1}^{N_\sigma} e^{i\theta_{\sigma,j}}.
\label{z separated}
\end{equation}
Defining local order parameters for each module as
\begin{equation}
    z_\sigma = \frac{1}{N_\sigma} \sum_{j=1}^{N_\sigma} e^{i\theta_{\sigma,j}} = r_\sigma e^{i\theta_\sigma}
\label{z individual}
\end{equation}
and $\rho_\sigma=N_\sigma/N$ where $N=\sum N_\sigma$ we can write $z$ in the compact form
\begin{equation}
    z = \sum_{\sigma=1}^{s} \rho_\sigma z_\sigma.
\label{z rho}
\end{equation}
Multiplying this equation by its complex conjugate gives
\begin{equation}
    r^2 = \sum_{\sigma=1}^{s} \sum_{\sigma'=1}^{s} \rho_\sigma \rho_{\sigma'} r_\sigma r_{\sigma'} e^{i(\theta_\sigma-\theta_{\sigma'})}.
\label{r^2 long}
\end{equation}

Rearranging the terms and defining the phase offset between modules $\sigma$ and $\sigma'$ by $\phi_{\sigma\sigma'}=\theta_{\sigma}-\theta_{\sigma'}$ we arrive at the final expression for the global order parameter
\begin{equation}
    r^2 = \sum_{\sigma=1}^{s} \rho_\sigma^2 r_\sigma^2+2\sum_{\substack{\sigma,\sigma'=1 \\ \sigma'>\sigma}}^{s} \rho_\sigma\rho_{\sigma'}r_\sigma r_{\sigma'}\cos{\phi_{\sigma\sigma'}}.
\label{z rho phi}
\end{equation}

For the coarse-grained network with $s$ nodes the order parameter is given by a similar expression, but with $r_\sigma=1$.  Also, by Eq. \eqref{multi pop II/ku sigma indexed}, in the coarse-graining regime the offset $\phi_{\sigma\sigma'}$ between modules is the same offset obtained for the system of $s$ oscillators. Therefore, the expression
\begin{equation}
    r_+^2 = \sum_{\sigma=1}^{s} \rho_\sigma^2 +2\sum_{\substack{\sigma,\sigma'=1 \\ \sigma'>\sigma}}^{s} \rho_\sigma\rho_{\sigma'} \cos{\phi_{\sigma\sigma'}}.
\label{z rho phi_up}
\end{equation}
defines an upper bound for the order parameter of the full network, representing a situation where all modules are in perfect internal synchrony and the system behaves exactly as $s$ oscillators,  weighted by size of modules they represent. We can determine the offset analytically for the case of two modules; for three and more modules we appeal to numerical solutions.

In the next sections we will use $r_\sigma$, the inner modular synchrony, as control parameters of our coarse-graining procedure. If $r_\sigma <<1$, modules loose  cohesion and the coarse-graining becomes less accurate. In order to avoid that, we shall assume that the intensity of the internal coupling $\lambda_\text{in}$ is large enough so that modules have a minimum internal synchrony $q_\sigma$. Replacing $r_\sigma \to q_\sigma$ gives an estimate for $r$ that takes into account both the relative module sizes and their level of synchronization: 
\begin{equation}
    r_-^2 = \sum_{\sigma=1}^{s} \rho_\sigma^2 q_\sigma^2+2\sum_{\substack{\sigma,\sigma'=1 \\ \sigma'>\sigma}}^{s} \rho_\sigma\rho_{\sigma'}q_\sigma q_{\sigma'}\cos{\phi_{\sigma\sigma'}}.
\label{z rho phi_low}
\end{equation}
If the coarse-graining process is successful, the order parameter $r$ for the full network should satisfy $r_- < r < r_+$ where $r_\pm$ are computed with the reduced network with $s$ nodes. However, as we shall see in the next section and in section \ref{real}, some modules might not reach the desired level of synchrony. In that case $q_\sigma$ should be replaced by the actual value of $r_\sigma$ reached by the module.

\section{Two modules}
\label{two}

We start by briefly describing the dynamics of $N=2$ Kuramoto oscillators. In this case, Eq. \eqref{ku} can be rewritten in terms of the oscillator's phase difference $\phi\equiv\theta_1-\theta_2$ as 
\begin{equation}
    \dot{\phi} = \omega - \lambda \sin{\phi}
\label{2 osc phi}
\end{equation}
where $\omega\equiv\omega_1-\omega_2$ is the oscillator's frequency offset.  For $\lambda > \omega$, Eq. \eqref{2 osc phi} has a stable fixed point at $\phi^\ast=\arcsin{(\omega/\lambda)}=\sqrt{1-(\omega/\lambda)^2}$. It is straightforward to derive from Eq. \eqref{z} that the order parameter takes the form $r^2=(1+\cos{\phi})/2$. This result is independent of the synchronization status of the system. So when the system is not synchronized we can write a mean and a deviation of $r$. The results are $\text{E}[r]=2/\pi$ and $\text{Var}[r]=1/2-4/\pi^2$. This closes our short discussion on two oscillators systems.

\begin{figure}[htb!]
    \centering
    \includegraphics[width=1.0\textwidth]{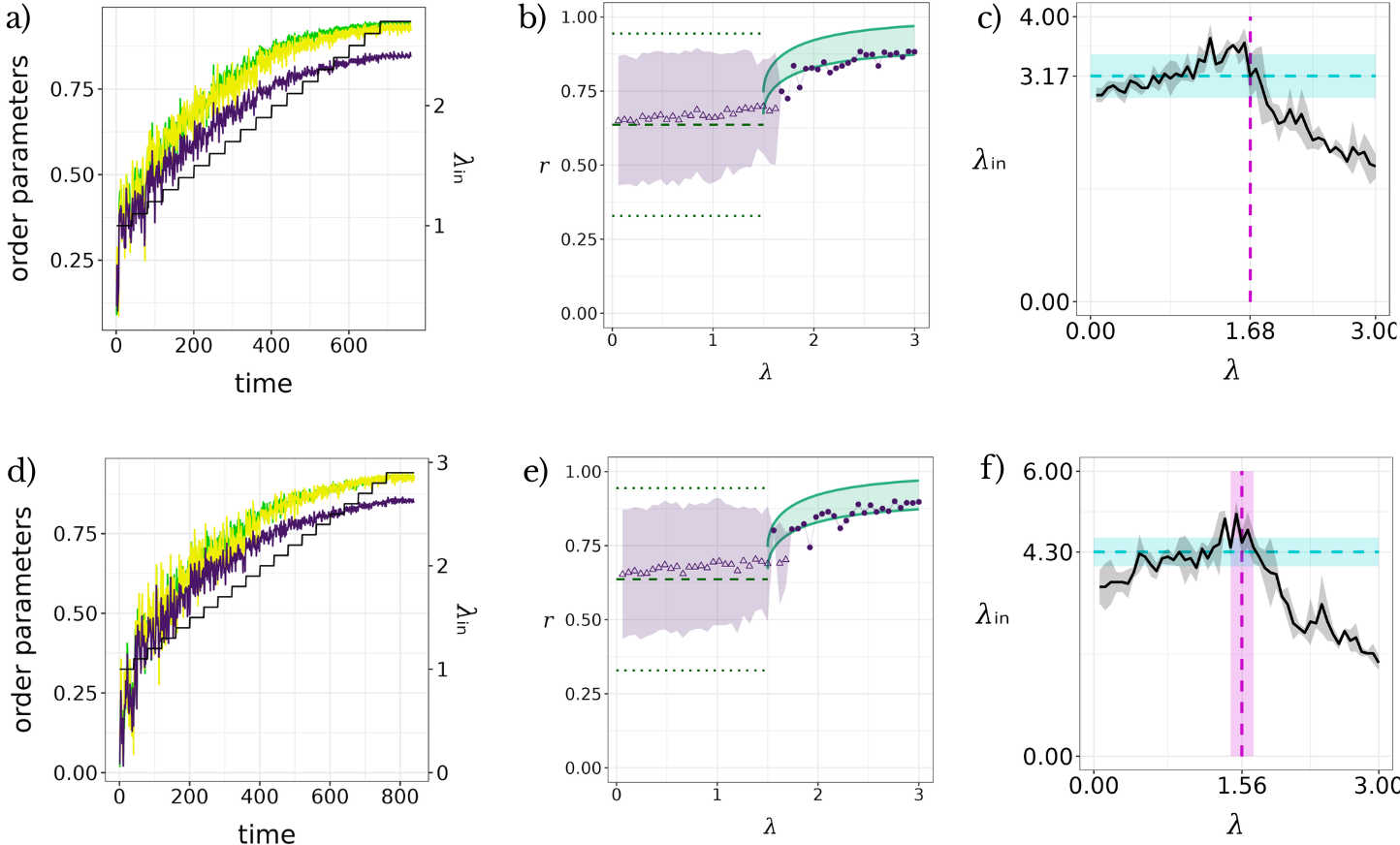}
    \caption{Simulation of a two-module network with $N_1=200$, $N_2=100$ and Gaussian frequency distributions with $\Delta=1$ and $(\omega_1,\omega_2)=(1.5,0)$. Panels (a)-(c) show results for fully connected modules, with $A_{ij}^{\sigma\sigma}=1$, and inter-module nodes connected with probability $p=0.1$. Panels (d)-(f) correspond to an Erdös-Rényi network, where every pair of nodes is connected with probability $p=0.1$, irrespective of their location in module 1 or 2. Panels (a) and (d) show the order parameters $r_1$, $r_2$ and $r$ as a function of the time in green, yellow and purple, respectively, for $\lambda=2$. The black ladder function shows how $\lambda_\text{in}$ was increased over time to achieve high inter-module synchronization, right $y$-axis. Panels (b) and (e) show the global order parameter as a function of inter-module coupling $\lambda$ with its standard deviation; triangles represent $\sigma>0.01$ and circles otherwise. Green solid curves represent Eqs. \eqref{z rho phi_up} and \eqref{z rho phi_low}. Theoretical mean and standard deviation for the asynchronized region are shown with green dashed lines. Panels (c) and (f) show the value of internal coupling for each external coupling such that desired internal synchrony $r_\sigma \geq q_\sigma$ was achieved. Magenta and turquoise lines marks the average of critical couplings over 3 independent simulations. A set of such pairs for various $\omega$ and $p$ are shown in Figure \ref{Fig2}. Simulations in panels (b) and (e) where performed with $\lambda_\text{in}$ at the critical value.}
    \label{Fig1}
\end{figure}

Now we proceed to simulate two modules systems. Since typical modules do not have the same size we will show simulations for the case where $N_1=200$ and $N_2=100$. In our first numerical experiment we assume that both modules are fully connected, i.e., $A^{\sigma \sigma}_{ij}=1$ if $i\neq j$ for $\sigma=1,2$. The frequency distributions $g(\omega)$ are Gaussians of width $\Delta=1$ and centered at $\omega_1=1.5$ and $\omega_2=0$, for modules 1 and 2 respectively. The synchrony threshold for both modules is set to $q_\sigma=0.9$. Figure \ref{Fig1} (a) illustrates the procedure to reach $q_\sigma$ on a network where the inter-module connection probability was set to $p=0.1$ and $\lambda=2$ . We started by simulating Eq. \eqref{multi pop II/ku} with internal coupling $\lambda_{\text{in}}=1$ for a fixed time interval of $\Delta t=40$. This assures the system went through the transient and stabilized. If, after this interval, $r_\sigma < q_\sigma$, we increase $\lambda_{\text{in}}$ by $0.1$ and continue the simulation for another $\Delta t$. We repeat the process until $r_1$ (green) and $r_2$ (yellow) surpass the boundary of $q_\sigma$ synchrony during a whole $\Delta t$. This determines the minimum $\lambda_{\text{in}}$ for the validity of the coarse graining procedure for given $\lambda$ and $p$.

The purple curve in Figures \ref{Fig1} (a) represents the global order parameter. After reaching minimum $\lambda_{\text{in}}$ for internal synchrony $q_\sigma$ we let the system evolve for a final time interval. This allows us to calculate the average and standard deviation of $r$ and, consequently, classify the state of the full system as synchronized or not. Note the stable value of $r$, indicating small standard deviation and synchrony.

Calculating the order parameter $\langle r \rangle$ and standard deviation $\sigma_r$ for a set of values of $\lambda$, instead of just the one presented in Figure \ref{Fig1} (a), we get panel (b), showing $\langle r \rangle$ with one unit of standard deviation after reaching minimum $\lambda_{\text{in}}$ (purple symbol and shaded area). For small $\lambda$, $r$ fluctuates and $\sigma_r$ is large whereas for large $\lambda$ the order parameter converges to a stationary value with small standard deviation. We define the phase transition to global synchrony as the lowest $\lambda$ such that $\sigma_r$ is lower than the threshold of $0.01$. In the figure we use triangles for non-synchronized states and dots for synchronized states. The green curves shown in panel (b) are related to a two oscillators system. The left horizontal lines are the mean and variance of asynchronous states of two oscillators, as discussed in the beginning of this section. Note the similarity with the simulated system. The right green functions show the upper bound $r_+$, Eq.(\ref{z rho phi_up}) and $r_-$, Eq.(\ref{z rho phi_low}) for the case of two modules. It is impossible for the modules to synchronize at values above the upper boundary because it is a barrier of perfect oscillators which our modules would only asymptomatically reach in particular cases when $\lambda_{\text{in}}\rightarrow\infty$. Nonetheless, Eq.(\ref{z rho phi_low}) for two modules
\begin{equation}
    r_- = \sqrt{\rho_1^2q_1^2+\rho_2^2q_2^2+2\rho_1\rho_2q_1q_2\sqrt{1-\left(\frac{\omega}{\lambda}\right)^2}}
\label{r rho w lambda 2mod}
\end{equation}
represents a good fit for the global synchrony of the system. 

Recall that for each $\lambda$ we had to determine a minimum $\lambda_{\text{in}}$ that fulfills the condition $r_\sigma\geq q_\sigma$. These values are shown in Figure \ref{Fig1} (c). We obtain a tent shaped curve that creates two regions: above the curve the coarse graining process is valid, and bellow the curve, it is not. Global phase transition is marked by the magenta line, while turquoise indicates the respective $\lambda_{\text{in}}$ for this globally synchronized state. The onset of synchronization is marked by the intersection of the magenta and turquoise lines.

Observe that before the phase transition, to satisfy the condition $r_\sigma\geq q_\sigma$ requires larger $\lambda_{\text{in}}$ as we increase $\lambda$, i.e, the strength of connections among modules disrupts the cohesion within the modules. On the other hand, after transitioning to global synchrony the parameter $\lambda$ helps bringing modules close together and consequently, demands less from the inner-module coupling strength.

In the previous experiment, both modules were considered to be fully connected. In order to relax this condition and study how the density of connections affect the coarse-graining, we show in panels (d)-(f) similar results for random modules, where internal nodes are also connected with probability $p=0.1$. Since we are setting the connection probability $p$ to be the same for every pair of nodes, the network falls into the Erdös-Rényi class. Modules are distinguished only by different connection strengths, $\lambda_\text{in}$ or $\lambda$.

\begin{figure}[htb!]
    \centering
    \includegraphics[width=1.0\textwidth]{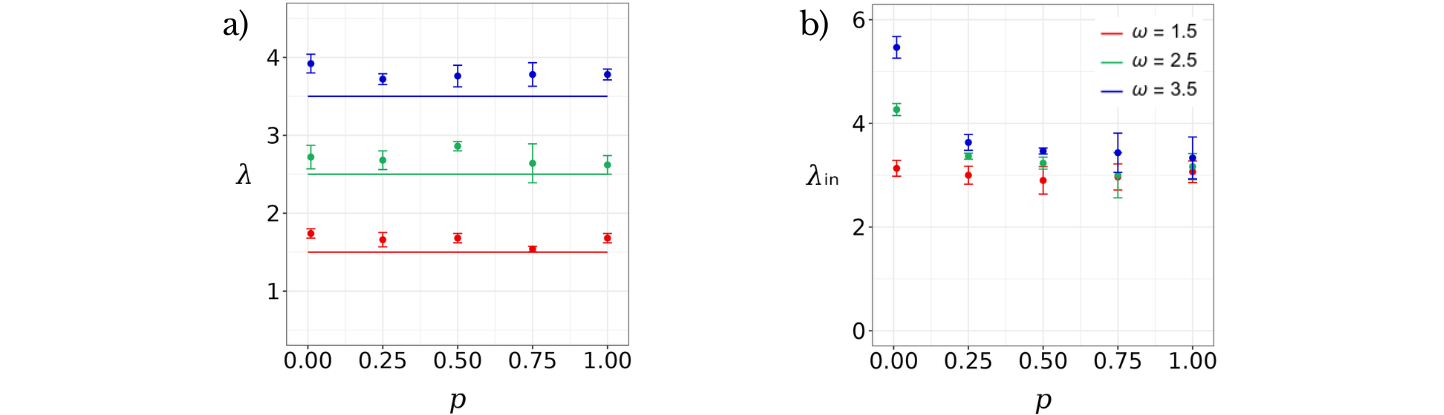}
    \caption{Critical inter-module (a) and inner-module (b) coupling strengths for fully connected modules as a function of coupling probabilities $p=0.01;0.25;0.5;0.75;1$ and frequencies $\omega=1.5;2.5;3.5$. Lines represent the corresponding 2 oscillator system. In all simulations $N_1=200$ and $N_2=100$; Gaussian frequency distributions with $\Delta=1$ and mean $(\omega_1,\omega_2)=(\omega,0)$. Error bars made over 3 simulations.}
    \label{Fig2}
\end{figure}

The results are very similar to the fully connected case, although requiring larger values of $\lambda_\text{in}$ to reach the same internal synchronization threshold of $q_\sigma=0.9$. The coarse-graining procedure is, therefore, largely independent of the network structure, as expected from the theory. However, we note that the validity of the process relies on the synchronizability of the modules, which is not always attainable. Ling et al. showed that for an Erdös-Rényi network the connectivity threshold for synchronization is $p\sim\log{(n)}/n$ which is about 0.03 for $n=200$ \cite{ling2019landscape,kassabov2022global}. Even though our simulated Erdös-Rényi network has different couplings to account for modularity, we confirmed that for small enough connection probability ($p=0.01$) modules do not synchronize regardless of $\lambda_\text{in}$ (simulations not shown). In section \ref{real} we will encounter this situation again.

The onset of synchronization indicates both minimum $\lambda_{\text{in}}$ for $r_\sigma\geq q_\sigma$ and $\lambda$ for global synchrony. Figure \ref{Fig2} shows such $(\lambda_{\text{in}},\lambda)$ pairs for 5 different values of frequency offsets $\omega$ and 5 different inter-module connection probabilities $p$ for fully connected modules. Panel (a) shows that, for large enough $\lambda_{\text{in}}$, the global phase transition happens at about $\lambda=\omega$ which is precisely when the phase transition of a 2 oscillator system takes place, in accordance with the coarse-graining process. More interesting, the transition point does not depend upon the number of connections between modules, as expected from Eq. \eqref{multi pop II/ku sigma indexed}. Panel (b) shows that immediately after the transition to the globally synchronized state, the minimum $\lambda_{\text{in}}$ that would assure sufficient synchrony for the coarse-graining process decreases non linearly with the amount of connections between the modules, exposing the importance of network structure in keeping inner-module cohesion.

\section{Three modules}
\label{three}

We now extend the results of the previous section to systems with three modules, as there are still some analytical tools available to represent the states of three Kuramoto oscillators. We divide the work into two parts. First, we find an approximation for the global phase transition in parameter space, Eq. \eqref{transicao fase 3 osc}. Secondly, we write Eq. \eqref{z rho phi} for 3 modules to find upper and lower bounds within the synchronization region.

\subsection{Three oscillators}

For 3 oscillators the Kuramoto model can be written in terms of the phase differences $\phi_{12}\equiv\theta_1-\theta_2$ and $\phi_{23}\equiv\theta_2-\theta_3$ as
\begin{subequations}
\label{3 osc phi}
    \begin{align}
        \dot{\phi}_{12} & = \Omega_1 -2\sin{(\phi_{12})}-\sin{(\phi_{12}+\phi_{23})}+\sin{(\phi_{23})}
        \label{3 osc phi/1} \\
        \dot{\phi}_{23} & = -\Omega_3 + \sin{(\phi_{12})}-\sin{(\phi_{12}+\phi_{23})}-2\sin{(\phi_{23})}
        \label{3 osc phi/2}
    \end{align}
\end{subequations}
where we rescaled time $t\rightarrow 3t/\lambda$, set $\omega_2=0$ (which is equivalent to a change in reference frame) and defined $\Omega_1\equiv3\omega_1/\lambda$ and $\Omega_3\equiv3\omega_3/\lambda$. These equations are similar to Adler's equation $\dot\phi = \Omega-\sin\phi$, but in 2 dimensions. Note that every point in the $\phi_{12}\times\phi_{23}$ plane is a fixed point for some set of parameters $\{\Omega_1,\Omega_3\}$, nonetheless, only some of these fixed points will be stable. Fixed points are solutions of
\begin{subequations}
\label{pts fixos w1 w3}
    \begin{align}
        \Omega_1 = 2\sin{(\phi_{12})}+\sin{(\phi_{12}+\phi_{23})}-\sin{(\phi_{23})}
        \label{pts fixos w1 w3/1} \\
        \Omega_3 = \sin{(\phi_{12})}-\sin{(\phi_{12}+\phi_{23})}-2\sin{(\phi_{23})}.
        \label{pts fixos w1 w3/2}
    \end{align}
\end{subequations}

\begin{figure}[htb!]
    \centering
    \includegraphics[width=1.0\textwidth]{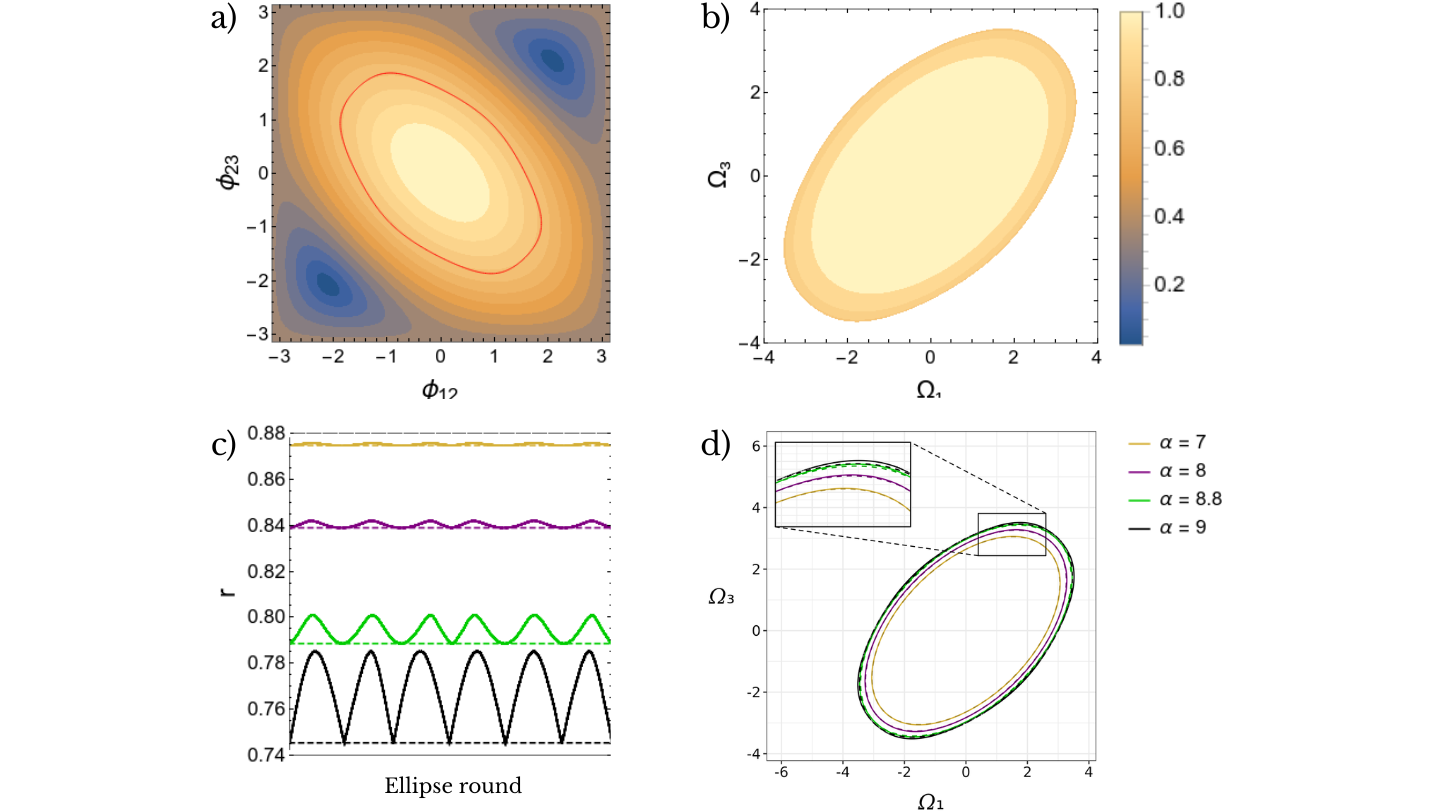}
    \caption{(a) Contour plot of Eq. \eqref{3 osc r of phi} in the phase space $\phi_{12}\times\phi_{23}$. The red curve divides stable (inside) from unstable (outside) points. (b) Stable points in the $\Omega_1\times\Omega_3$ plane according to Eq. \eqref{pts fixos w1 w3}. (c) $r$ values along the ellipses. Dashed lines are the empirical values from Eq. \eqref{3 osc r proposed w12 w23}. (d) Contrast between the ellipses (Eq.\eqref{3 osc ellipse regions w12w23}, dashed) and lines of constant $r$  (continuous).}
    \label{Fig3}
\end{figure}

Similar to two oscillators, the order parameter can be written in terms of phase differences as
\begin{equation}
    r = \frac{1}{3}\sqrt{3+2[\cos{\phi_{12}}+\cos{(\phi_{12}+\phi_{23})}+\cos{\phi_{23}}]}.
\label{3 osc r of phi}
\end{equation}
Figure \ref{Fig3} (a) shows the contour plot of Eq. \eqref{3 osc r of phi} in the phase space $\phi_{12}\times\phi_{23}$. To determine the stability of these fixed points we calculated the Jacobian of Eqs. \eqref{3 osc phi} and looked at the largest real part of the pair of eigenvalues: when it is zero we are at the critical line, shown as the red curve in Fig.\ref{Fig3} (a) . The curve divides the phase space into two regions: inside where the system displays stable synchrony, and outside, where motion is periodic. More elucidating, however, is to understand synchronization in the parameter space $\Omega_1\times\Omega_3$. Figure \ref{Fig3} (b) shows the stable $r$ values from panel (a) reparameterized according to Eq. \eqref{pts fixos w1 w3}. It is interesting to note that, regardless of being displayed in the phase space or the parameter space, contours of constant synchronization resemble ellipses. We proceed to finding approximations for such contours.

In a simpler version of the problem, if any pair of the 3 oscillators is identical, stable solutions will imply equal phases for such pair. This means that either $\phi_{12}$, $\phi_{23}$ or $\phi_{12}+\phi_{23}$ is zero depending on the pair that is identical. If the remaining phase difference also reaches a fixed point, we can determine it through system \eqref{3 osc phi} and replace it into Eq. \eqref{3 osc r of phi} to write $r$ in terms of $\Omega$, the frequency difference between the non-identical oscillators, as 
\begin{equation}
    r = \frac{1}{3}\sqrt{5+4\sqrt{1-\frac{\Omega^2}{9}}}.
\label{3 osc r w12=0 (w23=0)}
\end{equation}

Eq. \eqref{3 osc r w12=0 (w23=0)} gives the synchronization order parameter when 2 oscillators are identical, i.e, along the axes $\Omega_1=0$ $(\Omega=\Omega_3)$, $\Omega_3=0$ $(\Omega=\Omega_1)$ and $\Omega_1=\Omega_3$ $(\Omega=\Omega_1=\Omega_3)$ of Figure \ref{Fig3} (b). Thus, we empirically propose the constant synchronization contours to be ellipses described by
\begin{equation}
    \Omega_1^2-\Omega_1\Omega_3+\Omega_3^2=\alpha
\label{3 osc ellipse regions w12w23}
\end{equation}
and we modify Eq. \eqref{3 osc r w12=0 (w23=0)} into 
\begin{equation}
    r = \frac{1}{3}\sqrt{5+4\sqrt{1-\frac{\alpha}{9}}}
\label{3 osc r proposed w12 w23}
\end{equation}
with $\alpha \leq 9$ for the general case. This approximation indicates a first order phase transition from periodic motion to synchronization at $\alpha\equiv \alpha_1=9$ or, in terms of the original parameters $\omega_1$, $\omega_3$ and $\lambda$, at $\lambda_\text{c}^2=\omega_1^2-\omega_1\omega_3+\omega_3^2$.

Figures \ref{Fig3} (c)-(d) show the accuracy of this expression. The closer $(\Omega_1,\Omega_3)$ is to the origin, and consequently the lower the $\alpha$, the better the ellipse matches the real constant $r$ contour. As we move away from the origin, our proposed ellipses start to cut through several real fixed $r$ contour lines. In that case, what we observe is that Eq. \eqref{3 osc r proposed w12 w23} is returning the lowest synchronization value attainable over all the real contours that it is crossing.

In the inset of Figure \ref{Fig3} (d) we can easily see that there are constant $r$ values that fall out of the theoretical ellipses, therefore, there should be a smaller ellipse that encompass all values. We numerically find that further increasing $\alpha$ up to $\alpha_2\equiv9.295$ causes all possible synchronization outcomes to tightly fall within the ellipse. In other words, depending on how we control our parameters $\{\Omega_1,\Omega_2\}$ the phase transition happens somewhere between $\alpha_1\leq\alpha\leq\alpha_2$, or in terms of the original parameters,
\begin{equation}
    \lambda^2\leq\omega_1^2-\omega_1\omega_3+\omega_3^2\leq1.0328\lambda^2.
\label{transicao fase 3 osc}
\end{equation}

This is an estimate for the phase transition of 3 oscillators. As we argued in previous sections, both the coarse-grained system and the equivalent three oscillators system have the same phase transition due to the form of Eq. \eqref{multi pop II/ku sigma indexed}. However, it is important to emphasize that so far we do not have an approximation for $r$ within the ellipse that takes into account modules sizes and synchronization thresholds.

\subsection{Coarse graining for three modules}

Writing Eq. \eqref{z rho phi} for 3 modules we obtain 
\begin{equation}
    r^2 = \rho_1^2r_1^2+\rho_2^2r_2^2+\rho_3^2r_3^2+2\rho_1\rho_2r_1r_2\cos{(\phi_{12})}+2\rho_1\rho_3r_1r_3\cos{(\phi_{12}+\phi_{23})}+2\rho_2\rho_3r_2r_3\cos{(\phi_{23})},
\label{r rho phi 3mod}
\end{equation}
from which we can write $r_+$ and $r_-$. Despite having an expression for the order parameter in terms of the stable phase differences we do not have explicit solutions for such phases, only implicit ones given by the set \eqref{pts fixos w1 w3}. Consequently, determining a theoretical closed expressions as we did in Eq. \eqref{r rho w lambda 2mod} is not possible. Our approach for 3 modules thus focuses on writing $r_+$ and $r_-$ via numerically solving system \eqref{pts fixos w1 w3} to substitute into Eq. \eqref{r rho phi 3mod}. Note that the particular case of $\rho_i=1/3$ and $r_i=1$ throws us back to the 3 oscillators order parameter, Eq. \eqref{3 osc r of phi}.

\begin{figure}[htb!]
    \centering
    \includegraphics[width=1.0\textwidth]{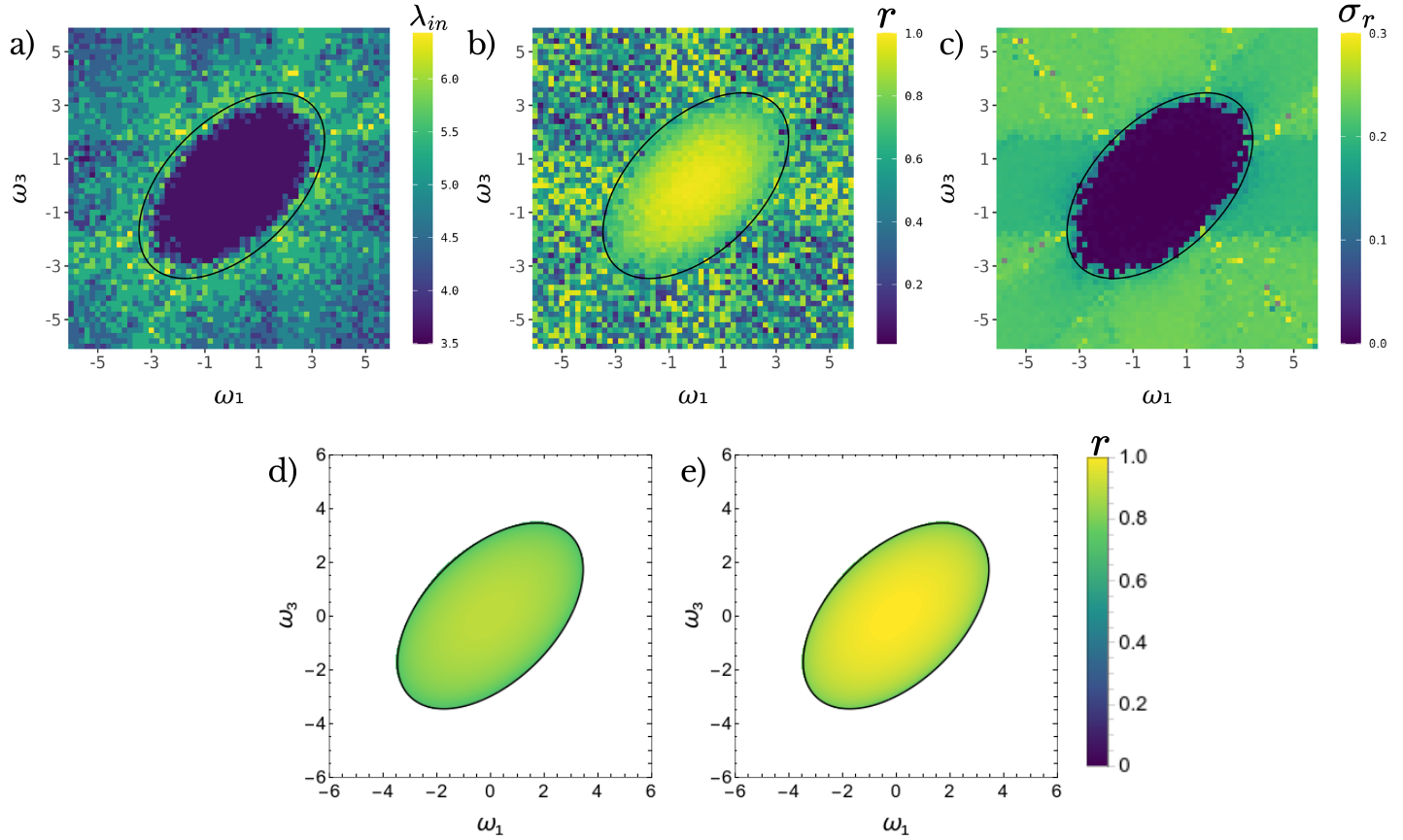}
    \caption{Simulation of Eq. \eqref{multi pop II/ku} with 3 modules. Panels (a)-(c) show final values of $\lambda_{\text{in}}$, global order parameter $r$ and standard deviation $\sigma_r$ in the frequency parameter space. Panels (d) and (e) show numerical estimates of $r_+$ and $r_-$ taken from Eq. \eqref{r rho phi 3mod} with $r_\sigma=1$ and $r_\sigma=q_\sigma=0.9$, respectively. Phase differences were obtained numerically, solving system \eqref{pts fixos w1 w3}. In all panels the black ellipse is the lower limit of inequality \eqref{transicao fase 3 osc} for the phase transition. Simulation values: $N_1=90$, $N_2=100$, $N_3=110$, $p=0.01$ and $\lambda=3$. Natural frequencies are Gaussian distributed with $\Delta=1$ and mean $\omega_1$, $\omega_2=0$ and $\omega_3$.}
    \label{Fig4}
\end{figure}

We now proceed to simulate Eq. \eqref{multi pop II/ku} with three modules and, as before, we take modules of distinct sizes to be more realistic, namely $N_1=90$, $N_2=100$ and $N_3=110$. Throughout this section we will work with fully connected modules, $p=0.01$ and $\lambda=3$ so that $\Omega_i=\omega_i$. Again, we set the frequency distributions to be Gaussians of width $\Delta=1$ centered at $\omega_1$, $\omega_2=0$ and $\omega_3$. The minimum synchrony of each module is fixed at $q_\sigma=0.9$.

Figure \ref{Fig4} (a) shows the final value of $\lambda_\text{in}$. We see the same behavior observed in the case of two modules: approaching the synchronization regime from a asynchronous state increases the internal coupling required to preserve $r_\sigma\geq q_\sigma$. This might seem an odd, considering we are decreasing the absolute frequency value and consequently expect it to be easier to keep cohesive modules. However, given that there are few connections between modules, the transfer of information from one to the others is harder when their frequencies are further apart, even though the absolute value is smaller. 

Figures \ref{Fig4} (b)-(c) show the global order parameter and its standard deviation, respectively. The black curve shows the lower elliptic bound given by Eq.\eqref{transicao fase 3 osc}. We observe that the synchronization region corresponds to the expected behavior for three oscillators and exhibit all distinct behaviors, full and partial phase locking and asynchrony. Panels (d) and (e) show the numerical estimates of the synchronization bounds $r_+$ and $r_-$ taken from Eq. \eqref{r rho phi 3mod}, respectively. The phase differences were obtained by numerically solving system \eqref{pts fixos w1 w3}. Note that the simulated value of $r$ appears to fall in between both, as expected.

\section{Real Networks}
\label{real}

We have seen so far that, given high enough $\lambda_\text{in}$, the coarse-graining procedure recreates the synchronization properties of the reduced system in artificial modular networks with fully connected and Erdös-Rényi  modules (shown only for the case of two modules). These networks, however, were generated in such a way as to exhibit very strong modular structure that does not resemble real world systems. In what follows we test our results on two well known networks, Zachary's Karate club social network \cite{zachary1977information} and the \textit{C. Elegans} gap junctions neural network \cite{white1986structure}. Both of these networks can be modularized using standard procedures. What distinguishes them from the artificial networks constructed in this paper are structural properties such as not having full number of connections within modules and non random connections between modules per se. 

\begin{figure}[htb!]
    \centering
    \includegraphics[width=1.0\textwidth]{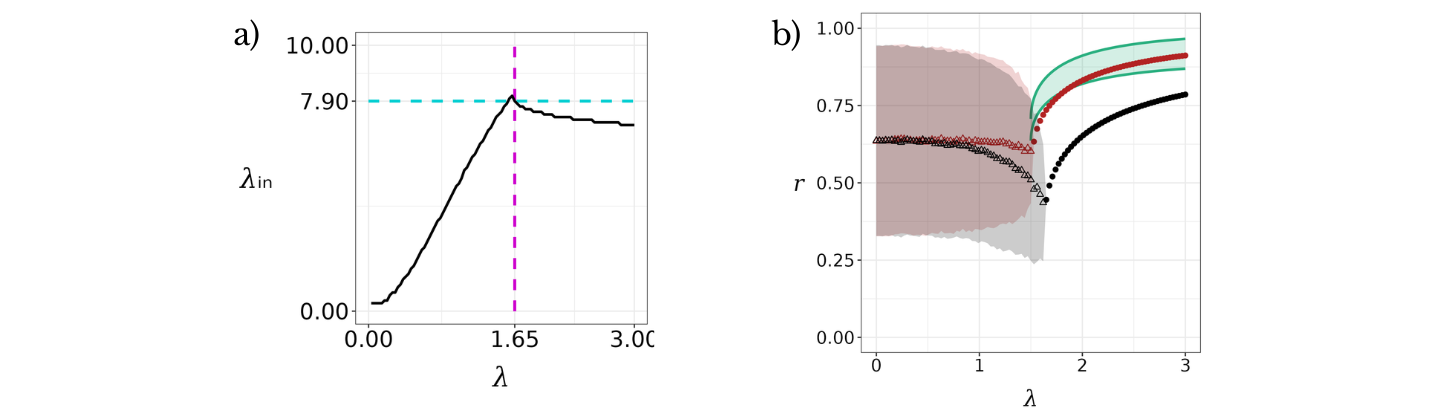}
    \caption{Dynamics of Eq. \eqref{multi pop II/ku} for Zachary's Karate social network. (a): Final $\lambda_{\text{in}}$ and a function of $\lambda$. Magenta and turquoise lines marks the average of critical $\lambda$ and $\lambda_{\text{in}}$. (b): $\langle r \rangle$ with $\lambda_{\text{in}}=7.9$ (the critical value) in black and with $\lambda_{\text{in}}=20$ in red. Green solid curves represent the theoretical results, Eqs. \eqref{z rho phi_up} and \eqref{z rho phi_low}. Simulation values: $\Delta=0$, $\omega_1=1.5$ and $\omega_2=0.0$.}
    \label{Fig5}
\end{figure}

We start with the Karate network \cite{zachary1977information}, that has a total of 34 nodes with 78 connections. To modularize the network we used Girvan-Newman's method, splitting it in two modules of 17 nodes each \cite{girvan2002community}. Connections are divided into 32 and 35 for each module and 11 connecting the modules. As before, we run the dynamics given by Eq. \eqref{multi pop II/ku} on this network and slowly increase $\lambda_{\text{in}}$ until a minimum inner module synchrony of 90\% is reached. Figure \ref{Fig5} (a) shows the minimum internal coupling that keeps the modular synchrony above the threshold. Note that the global phase transition, expected at $\lambda=1.5$ given the strong bi-modular characteristic of Zachary's Karate network, ocurs at $\lambda=1.65$. Just as in the artificial network, increasing $\lambda$ for asynchronized states impairs modular synchrony, hence the continuous increase in $\lambda_{\text{in}}$ as $\lambda$ grows. On the other hand, after the phase transition, synchronized states act in cohesion, requiring lesser $\lambda_{\text{in}}$ as we increase $\lambda$. However, the slopes of increasing and decreasing values of $\lambda_{\text{in}}$ are quite different from what we found in section \ref{two}. This behavior reflects the quantitative differences between $\lambda_\text{in}$ values among the real and artificial networks, Figures \ref{Fig1}(c) and \ref{Fig5}(a). The information transfer among modules is harder in Karate's network, which has proportionally fewer connections among modules, 11 out of 289, corresponding to $p\approx0.04$, as opposed to $p=0.1$ for the artificial fully connected network. Consequently, Zachary's Karate network requires much lower $\lambda_\text{in}$ prior to the phase transition and higher after it.

The order parameter as a function of $\lambda$ is shown in Figure \ref{Fig5}(b). We fixed the internal coupling to the minimum value that assures modular synchrony, i.e., $\lambda_\text{in}=7.9$, corresponding to the peak in panel (a). The order parameter (black points), shows that synchronized states do not reach the $[r_+,r_-]$ region, in green. This is likely due to the low connectivity between modules which lead to large inter-modular phase difference. The global phase transition point is also slightly larger than the expected value of 1.5. We can counteract this structure effect by overshooting the internal coupling. The red curve of Figure \ref{Fig5} (b) shows the synchronization values when keeping $\lambda_{\text{in}}=20$ throughout the whole simulation. Forcing high inner-module coupling strength pushes both modules to highly synchronized states so that the system behaves almost exactly as two oscillators.

\begin{figure}[htb!]
    \centering
    \includegraphics[width=1.0\textwidth]{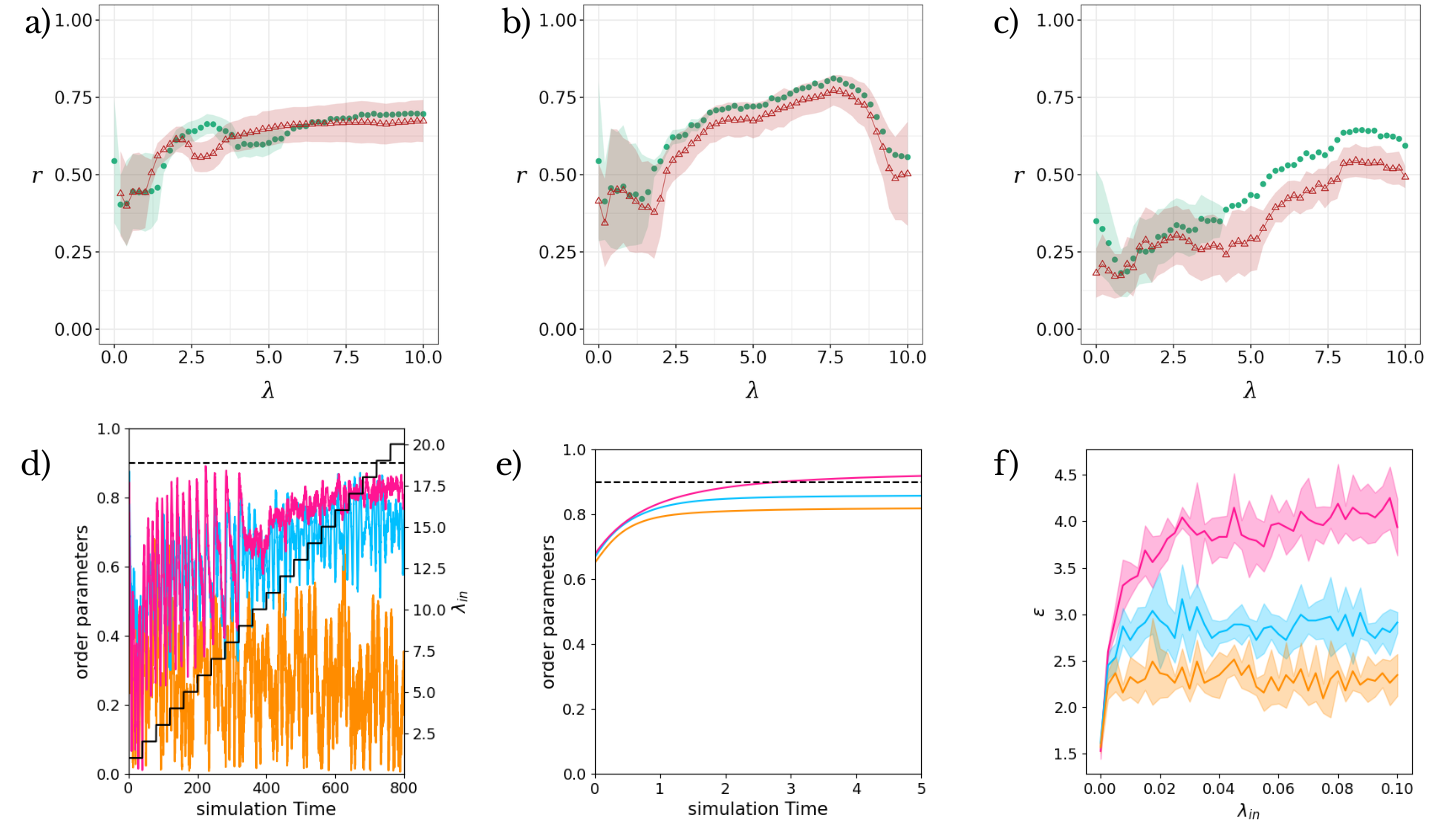}
    \caption{Dynamics of Eq. \eqref{multi pop II/ku} for \textit{C. Elegans} gap junctions network. (a), (b) and (c) show $\langle r \rangle$ for networks clustered into 3, 5 and 10 modules with $\lambda_{\text{in}}=20$ (red) and coarse grained prediction $r_-$ (green). Simulation values: $\Delta=0$, (d): $(\omega_1,\omega_2,\omega_3)=(0,1,6)$; (e): $(\omega_1, \omega_2, \omega_3, \omega_4, \omega_5)=(0,1,1,-2,7)$; (f): $(\omega_1,\omega_2,\omega_3,\omega_4,\omega_5,\omega_6,\omega_7,\omega_8,\omega_9,\omega_{10})=(0,1,-3.5,-2,2,0.6,0.4,1.2,1.5,-1.5)$. panels (d) and (e) show the coupled ($\lambda=2$, $\lambda_\text{in}=20$) and decoupled ($\lambda=0$, $\lambda_\text{in}=1$) dynamics of 3 modules for same initial phases. Panel (f) shows the maximum angular spread of initial phases $\epsilon$ such that the decoupled modules synchronize above 90\%.}
    \label{Fig6}
\end{figure}

As for our second example we study a reduced network extracted from the \textit{C. Elegans} gap junctions neural network containing 248 nodes and 511 connections \cite{OWP}. Using appropriate metrics it can be modularized into 3, 5 or 10 modules \cite{szalay2012moduland,moreira2020synchronization,moreira2019modular}. Note that regardless of number of modules, some end up being far greater than others, for instance, in the 3 modules arrangement we have 130 nodes in the largest module, 77 nodes in the intermediate and 41 nodes in the smallest. For 5 modules the number of nodes in each module are 120, 83, 34, 7 and 4 and with 10 modules the module sizes are 68, 33, 76, 11, 13, 7, 6, 8, 8 and 18. The modularity coefficient \cite{girvan2002community} for these partitions is $Q=0.44$ for 3 and 5 modules and $Q=0.55$ for 10 modules \cite{moreira2020synchronization}. Moreover, the network has about 1\% of the total number of possible connections, indicating that it is hard to make all modules have a minimum synchrony of 90\% or even lower. For that reason we made some simplifying assumptions to test if the coarse graining procedure could work at least under favorable conditions. First, we assumed that oscillators in each module are identical, i.e. $\Delta=0$. Second, given the small density of connections in some of the modules, we attributed initial phases for the oscillators in the interval $0 < \theta_{\sigma,i} < \epsilon$, where $\epsilon$ was typically around $\pi$. This condition guarantees that the initial conditions are within the basin of attraction of the synchronized state for each module. That, however, does not imply that all modules will reach the desired synchronization level, as they are coupled to the rest of the network. The values of natural frequencies for the modules where chosen in a rather arbitrary way (see caption of Figure \ref{Fig6} for specific values) to produce a non-trivial global order parameter. 

Before we compare exact and coarse-grained results for the \textit{C. Elegans} we emphasize that, although the network's backbone is always the same (same nodes and same connections) each partition of the network into modules leads to a different distribution of interaction strengths $\lambda_\text{in}$ (within modules) and $\lambda$ (between modules). Therefore each case responds to the Kuramoto dynamics differently.

Figure \ref{Fig6}(a)-(c) show in red the simulation of $\langle r\rangle$ for the 3 modularizations above, each accompanied by a simulation of the theoretical order parameter $r_-$ of the reduced system in green for comparison. For $r_-$ the phase differences were obtained by integrating the respective $s$ oscillators system and using for $q_\sigma$'s the mean values of $r_\sigma$ obtained from  the full network simulation as in Fig. \ref{Fig6}(d). In all cases the agreement is surprisingly good, despite the complex behavior and the relatively low value of global synchronization displayed by $\langle r\rangle$. Panels (d)-(f) show the details of the method for the case of 3 modules. In (d) we see how module 1 (red) reaches $q_1 \approx 0.8$, module 2, $q_2 \approx 0.7$, and module 3 does not synchronize even for large $\lambda_\text{in}$, attaining $q_3 \approx 0.2$ only. The green points in (a) were obtained using these values for the $q_\sigma$'s in Eq. \eqref{z rho phi_low}. Panel (e) shows the order parameters for the modules for $\lambda=0$ using the same initial conditions as in (d), indicating that they are synchronizable when independent, but that the interaction messes with their internal synchrony. Finally, panel (f) shows, also for $\lambda=0$, how the maximum spread $\epsilon$ of initial phases that allow convergence to the synchronized state increases with $\lambda_\text{in}$. We see, in particular, that module 3 only synchronizes if $\epsilon < 2.5$, even for identical oscillators and $\lambda=0$, indicating that the synchronized state is stable but with a not so large basin of attraction. A similar study was performed for the 5 and 10 modules partitions of the network, the corresponding $q_\sigma$'s were used to compute $r_-$.

\section{Discussion}
\label{conclusions}

In this work we proposed a coarse-graining method for modular networks where each module is replaced by a single effective node, drastically reducing the network size. The order parameter of the reduced network takes into account the relative size of the module represented by the effective node and its average degree of internal synchronization. To test the coarse-graining procedure we initially constructed artificial networks under two core assumptions: first, modules should be synchronizable, in the sense that for sufficiently large internal coupling $\lambda_\text{in}$ the local order parameter for the module should reach  values $r_\sigma$ of the order of $q_\sigma=0.9$. This ensures that the hypothesis of phase synchronization within the module is approximately true. Second, all intra-module connections are fixed to the same value $\lambda_\text{in}$, while inter-module connections take a value $\lambda$, which we vary to observe the transition to global synchronization. For synthetic networks with 2 and 3 modules the procedure works very well, as indicated in Figures \ref{Fig1} to \ref{Fig4}. We tested the approach for fully connected and randomly connected modules.

For the two real networks considered in section \ref{real} we relaxed the first condition, allowing $r_\sigma$ to be smaller than the desired value $q_\sigma=0.9$, as small modulus tend not to synchronize even for large internal coupling strength. That, however did not spoil the coarse graining, most likely because small modulus give small contributions to $r_\pm$. Taking this into account we showed that results for the \textit{C. Elegans} were surprisingly good for all partitions considered. We presume that this excellent agreement was possible because we considered identical oscillators in each module and chose initial phases not too far away from synchrony, facilitating synchronization for most of the modules. Further studies considering distributions of natural frequencies (with different average values) and random initial phases (as we did in sections \ref{two} and \ref{three} for artificial networks) would be important to understand the robustness of our results.

Given the experiments with artificial and real networks, we can summarize the coarse graining method according to the following recipe: (i) for a given network with $s$ modules,  fix a desired synchronization threshold $q_\sigma$; (ii) for each inter-module coupling $\lambda$, run the Kuramoto model with Eq. \eqref{multi pop II/ku} and vary the internal coupling $\lambda_\text{in}$ until $r_\sigma \geq q_\sigma$; (iii) from $\lambda_\text{in}$ as a function of $\lambda$ take the largest $\lambda_\text{in}$, which should correspond to $\lambda$ close to the phase transition of the network; (iv) if one or more modules do not reach the desired synchronization threshold $q_\sigma$, replace $q_\sigma$ by $\langle r_\sigma\rangle$ for these modules; (v) run the standard Kuramoto dynamics on the reduced network with $s$ nodes and compute the asymptotic phase differences $\phi_{\sigma \sigma'}$; (vi) compute $r_\pm$ using Eqs. \eqref{z rho phi_up} e \eqref{z rho phi_low} and compare with the order parameter of the original network.

It would also be interesting to compare our results with the Ott-Antonsen theory (OA) \cite{ott2008low} applied to modular networks. The OA ansatz reduces the dynamics of $s$ modules to $2s$ coupled differential equations, for module and phase of the order parameters computed for each module separately, from which the global order parameter can be calculated.  \cite{costa2024dynamicsmatrixcoupledkuramoto}. However it is important to keep in mind that such procedure works only for infinitely many oscillators and Lorentzian distribution of natural frequencies. Our approach differs from OA in the sense that we replace modules of the original network by effective oscillators and run the Kuramoto model on this smaller system. Although the method was inspired by EEG measurements, it does not require infinitely many oscillators per module nor specific distributions of natural frequencies. However, we do required modules to be "synchronizable". 

We should mention that we require prior knowledge of the network structure. In particular, we assume the network to be static in its composition and, therefore, the method does not accommodate merging or splitting of modules. Also, the fact that $\lambda_\text{in}$ is generally much larger than $\lambda$ in the vicinity of the phase transition ensures that inter modular dynamics evolve on a faster time scale than that of the full network, justifying the replacement of modules by effective nodes. 

We have seen that network properties like topology and link density are important for the proper application of the coarse graining procedure. Modules that are loosely connected do not synchronize and might affect the accuracy of the results. In the examples shown here for the \textit{C. Elegans}, such modules were small and did not compromise the quality of the results. It has been shown, for example, that to retain synchronization, some Laplacian eigenvalues should be preserved between the original and coarse-grained networks \cite{gfeller2008spectral}. Understanding the role of these properties in the coarse-graining process will help to unravel under what circumstances treating groups of nodes as single oscillators is a valid assumption. 

\begin{acknowledgments}

	It is a pleasure to thank J. A. Brum and Guilherme F. Arruda for helpful suggestions. Relevant codes and documentation are available at open access GitHub repository \cite{Codes}. This work was partially supported by FAPESP, grant 2021/14335-0 (MAMA) CAPES, grant 88887.706116/2022-00, and CNPq, grant 303814/2023-3 (MAMA). 
    
\end{acknowledgments}

\newpage
\bibliographystyle{ieeetr}

\end{document}